\documentstyle[aps,prl,multicol,epsfig]{revtex}
\begin{document}
\draft
\title{
Theory of superconductor-insulator transition
 in single Josephson junctions}
\author{
S.G. Chung
}
\address{
Max-Planck-Institut f\"{u}r Physik komplexer Systeme, N\"{o}thnizer
Str. 38, D-01187 Dresden, Germany
}

\maketitle

\begin{abstract}
A non-band theory is developed to describe the superconductor-insulator (SI)
transtition
in resistively shunted, single Josephson junctions. The $I-V$ 
characteristic is formulated by a Landauer-like formula
and evaluated by the path-integral transfer-matrix method.  The result is
consistent with the recent experiments at around 80 $mK$.  However, the
insulator phase shrinks with decreasing temperature indicating that
the single Josephson junction becomes all superconducting at absolute
zero temperature, as long as dissipation is present.

\end{abstract}



\begin{multicols}{2}

The resistively shunted single Josephson
junction(RSJ) at $T=0$ is described classically by the second order differential equation for the
phase difference between the two superconductors,

\begin{equation} \label{eq1}
C\frac{\hbar\ddot{\phi}}{2e}+
\frac{1}{R_{\rm T}}
\frac{\hbar\dot{\phi}}{2e}+
I_{0}{\rm sin}\phi
=I
\end{equation}
Here $C$ is the capacitance, $R_{T}$ is the
normal state resistance of the junction, $I_{0}$ is the critical current, and
$I$ is the bias current.
A mechanical analogy of a particle moving in a tilted washboard potential with
friction may be helpful. The $I-V$ characteristic resulting from (\ref{eq1}) was studied
by Stewart\cite{Stewart}.
The capacitance brings about hysteresis.  In the weak dissipation limit $R_{\rm T} 
\rightarrow  \infty $ current is entirely carried by Cooper
pairs at $V=0$ and up to the critical current.  This is the dc Josephson
effect.  For $I>I_0$, the $I-V$ curve switches to the Ohmic branch where current
is entirely carried by electrons, $V=RI$.  In the opposit limit $R_{\rm T}
\rightarrow 0$ hysteresis disappears and supercurrent at
$V=0$ is gradualy replaced by the normal current with increasing voltage.   The
thermal and quantum fluctuation is expected to affect only the regime
$I/I_{0}\ll R_{q}/R_{T}\equiv g$, where $R_{q}\equiv h/4e^2=6.45~K\Omega$ is
the quantum resistance.
Now for a typical charging energy of 1K, the critical current is
$I_{0}=2eE_{J}/\hbar=42E_{J}/E_{C}~nA$
whereas the parameter ranges associated with the SI transition are 
$g\lesssim 1,~E_{J}/E_{C}\lesssim 1$.
The range of current where fluctuation is important is thus up to a few $nA$.  
The focus of the present paper
is precisely in this current regime.  
It should be noted that if $g\gg 1$, the bias current could be as large as the
critical current while still staying in the regime where fluctuation is important. Indeed this case was studied
theoretically\cite{Zorin,Ingold} and experimentally\cite{Haviland} focusing on the negative
differential resistance as a signal of the Coulomb blockade of Cooper pair tunneling.

A microscopic description is to take into account the
 quasi-particle tunneling but with zero superconducting gap (resistively
 shunted) as a source of dissipation.
 The path-integral expression of the partition function is\cite{Ambegaokar},
\begin{equation} \label{eq2}
Z_{0}
=
\int{\cal D}\phi~{\rm exp}(-S/\hbar)
\end{equation}
\begin{eqnarray} \label{eq3}
S
&=& 
\int_0^{\beta\hbar} d\tau
\left[\frac{C}{2}\left(\frac{\hbar}{2e}\dot{\phi}
\right)^2 + U(\phi)\right]
\nonumber\\
& &
-\int_0^{\beta\hbar}\int_0^{\beta\hbar}
d\tau d\tau'\alpha(\tau-\tau') {\rm
cos}\left(\frac{\phi(\tau)-\phi(\tau')}{2}\right)
\end{eqnarray}
where
\begin{eqnarray} \label{eq4}
U(\phi)=
-E_{J}{\rm cos}\phi-\frac{I\hbar}{2e}\phi
\nonumber\\
\alpha(\tau)=
\frac{\hbar}{2\pi e^2R_{\rm T}}
\frac{(\pi/\beta\hbar)^2}
{{\rm sin}^2(\pi\tau/\beta\hbar)}
\end{eqnarray}
The single Josephson junction (JJ) has also been studied by a
 phenomenological model\cite{Gravert} which is just to replace the term ${\rm
cos}\left(\frac{\phi(\tau)-\phi(\tau')}{2}\right)$
in (\ref{eq3}) by $\phi(\tau)\phi(\tau')$ leading, upon stationally phase
approximation, to (\ref{eq1}) with retarded dissipation. 

The first question concerning the phase dynamics is whether it is
{\it coherent} or {\it incoherent}.  A band theory assumes a
coherent motion of phase over many local minima of the washboard potential.
 In band picture, the voltage calculation is
easy because it is proportional to the group velocity of the
quasi-charge\cite{Zorin}.  On the other hand, a non-band theory is appropriate for incoherent motion,
and the problem is reduced to the calculation of the escape rate of particle
out of a dissipative metastable well, the quantum Kramers rate.

Watanabe and Haviland\cite{Haviland} discussed their recent experiment 
on the Coulomb blockade at $g\gg 1$ in light of the band theory.
However, other experiments\cite{Hakonen,Yagi} have not found the negative differential
resistance in the SI transition
regime $g\lesssim 1$.  This indicates that the phase dynamics near the SI transition is dominated by
{\it incoherent} tunneling.  Indeed a typical experimental setup is, 
$E_{C}=1~K,~E_{J}/E_{C}
=0.1\sim20,~T/E_{C}=0.1\sim1$ and 
$g\lesssim 1$.  And most importantly, the minimum
voltage experimentally observable is of order $0.5\mu V$\cite{Hakonen}
which translates to
the temperature $ 10~mK$.  So when the energy splitting due to tunneling
exceeds this value, the corresponding state is experimentally recognized as insulating.
However, $10~mK$ is much smaller than the potential barrier 
  $E_{J}$ at the SI
transition.  Near the SI transition at $g=0$, therefore, the effect of many wells is expected
unimportant.  Switching on dissipation, the SI
transition shifts to a smaller critical 
$E_{J}/E_{C}$, but then
dissipation will suppress the effect of incresed tunneling strength.  One thus
expects that the phase dynamics essentially at a single well would determine
the SI transition.

In this paper we propose a non-band theory for the SI transition\cite{Praha}.
Two methods have been employed to calculate the quantum Kramers rate 
along this line.  One is the imaginary free energy method
with an instanton technique\cite{Gravert,Weiss}.  The other is the quntum
transition state (QTS) theory long known in chemistry\cite{Miller,Hanggi}.  Here
we develop a new approach combining a Landauer-like formula
for the quantum Kramers rate\cite{Chung1} and the cluster transfer matrix (TM) method\cite{Chung2}.
The Landauer formula for electrical conductance has proven to be quite useful
in mesoscopic systems.  In the quantum Kramers rate problem, there is formally
exact formula\cite{Yamamoto} analogous to the Green-Kubo-Nakano formula for electrical
conductivity.  The corresponding Landauer-like formula is the quantum
transient state (QTS) theory which tries to calculate the
Kramers rate by summing up all the
probabilities of finding particle at the top of the barrier which have
positive, outgoing momentum. 
The QTS theory, however, has often been exercised with a semiclassical
approximation\cite{Hanggi,Mel} and its drawbacks do not show up this way.  
As soon as one tries to exactly evaluate
the path-integrals, however, 
one encounters difficultiess.  First, the
contribution to the escape rate is no longer local.  The phase motion 
at different places than at the barrier top
contributes.  Moreover, the non-local contribution could become negative, as
discussed in\cite{Chung1}. This undesirable 
nonlocality and negativity
originates from the Weyle quantization procedure\cite{Weiss,Miller}.
The original idea
to collect all the outgoing, positive contribution {\it at} the barrier site,
is thus spoiled. 

We have recently proposed a Landauer-like formula for the quantum Kramers
rate. Applied to the voltage calculation here, it reads\cite{Chung1}
\begin{eqnarray} \label{eq5}
\frac{V}{e/2C}
=
\left[1-{\rm exp}(-\pi I\hbar\beta/e)\right]\times\frac{1}{2}
\sqrt{
\langle\delta(\phi-\phi_{0})P^2/L\rangle
}
\end{eqnarray}
where $P\equiv-\partial_\phi$, $\phi_0$ is the barrier top in the potential
$U(\phi)$, $L$ is the system size, and $<...>$ is a thermal average. The first
factor in (5) takes into account the backward flux by detailed balance,
whereas the second term is esentially square root of average of the kinetic
energy, in mechanical analogy, at the barrier top with factor 1/2 taking into
account only the right-going contribution.  In contrast, the QTS formula is
obtained from (5) by replacing
$\sqrt{
\langle\delta(\phi-\phi_{0})P^2/L\rangle}
\to
\langle\delta(\phi-\phi_{0})|P|\rangle
$.  The difference between the formula (5) and the QTS theory is thus a
fluctuation which is precisely the undesirable non-local, possibly negative
contributions in the QTS theory described in the above.
After some manipulations, one can express the average in (\ref{eq5}) as
\begin{eqnarray} \label{eq6}
\langle\delta(\phi-\phi_{0})P^2\rangle
=
Z/Z_0\\
Z
=
-\frac{1}{4}\partial_{\phi}^{2}W(\phi,2\phi_{0}-\phi)
\arrowvert_{\phi~=~\phi_{0}}\\
Z_{0}
=
\int d\phi W(\phi,\phi)
\end{eqnarray}
in terms of the path integral
\begin{equation} \label{eq9}
W(\phi,\phi')
=
\int_{\phi\to\phi'}{\cal D}
\tilde{\phi}~{\rm exp}(-S/\hbar)
\end{equation}
where $\tilde{\phi}(0)=\phi$ and $\tilde{\phi}(\beta\hbar)=\phi'$.

The path-integral (\ref{eq9}) with the action $S$ given by (\ref{eq3}) can be
evaluated precisely by the cluster TM method\cite{Chung2}.  In the present problem, however, the dimensionless
conductance is at $g\lesssim1$,
and from the study of the single electron box which has a similar action as
(\ref{eq3}), this regime of $g$ can be accurately handled by the 1-cluster TM
method as follows.  First divide the interval $(0, \beta\hbar)$ into $N$
segments of size $\Delta=\beta\hbar/N$.  Denoting the discrete phase points as
$\phi_{i},i=1,2,...,N+1$, we have
\begin{equation} \label{eq10}
{\rm exp}(-S/\hbar)
\equiv
\prod_{i=1}^N K(\phi_{i},\phi_{i+1})\times {\rm Rm} 
\end{equation}
where the TM operator $K$ is defined by
\begin{eqnarray} \label{eq11}
K
&=&
{\rm exp}\left[
-\frac{1}{16\Delta}(\phi_{i}-\phi_{i+1})^2
+J\Delta({\rm cos}\phi_{i}+I\phi_{i})
\right.
\nonumber\\
& &
\left.
-2g\frac{\Delta^{2}T^{2}}{{\rm sin}^{2}(\pi T\Delta)}\left
(1-{\rm cos}\frac{\phi_{i}-\phi_{i+1}}{2}\right)
\right]
\end{eqnarray}
with dimensionless parameters $J\equiv E_{J}/E_{C},~T\equiv k_{B}T/E_{C},~
I\equiv I\hbar/(2eJ)$
and
\begin{eqnarray} \label{eq12}
{\rm Rm}
\equiv
{\rm exp}\left[
-2gT^2\Delta^2\sum_{j-i\ge 2}^{N}
\frac
{1-\langle {\rm cos}\frac{\phi_{j}-\phi_{i}}{2}\rangle}
{{\rm sin}^2 \left(\pi T\Delta (j-i)\right)}
\right]
\end{eqnarray}
where the cumulant approximation was used which is precise
for $g\lesssim 1$.  Second write
\begin{equation} \label{eq13}
\int_{\phi\to\phi'}^{}{\cal D}
\tilde{\phi}
=\prod_{i=1}^{N+1}\int d\phi_{i}
\delta\left(\phi_{1}-\phi\right)
\delta\left(\phi_{N+1}-\phi'\right)
\end{equation}
The first $\delta$-function can be written in terms of a complete set
\{$\psi_{p}$\} as
\begin{equation} \label{eq14}
\delta\left(\phi_{1}-\phi\right)
=
\sum_{p}\psi_{p}^*(\phi)\psi_{p}(\phi_{1})
\end{equation}
Now one can choose the complete set such that it satisfies the TM equation,
\begin{equation} \label{eq15}
\int_{\phi_{a}}^{\phi_{b}}d\phi K(\phi,\phi')\psi_{p}(\phi)
=
\lambda_{p}\psi_{p}(\phi')
\end{equation}
Repeatedly using the TM equation, we arrive at
\begin{equation} \label{eq16}
W(\phi,\phi')
=
{\rm Rm}\cdot \sum_{p} \lambda_{p}^N \psi_{p}^* (\phi)\psi_{p}(\phi')
\end{equation}
The correlation function $\langle {\rm cos}\frac{\phi_{j}-\phi_{i}}{2}\rangle$
in (\ref{eq12})
can be calculated likewise.

There are three parameters $T, g$ and $E_{J}/E_{C}$.  In addition the present
method contains two more parameters $\Delta$ and $J_{R}$.  The latter defines
the phase space; $\phi_{a}=-\pi-\phi_{0}-\pi/J_{R}$, $
\phi_{b}=\pi-\phi_{0}+\pi/J_{R}$.  For numerical results obtained, we
have checked that the results are affected a couple of \% for the two different $J_{R}=2,3$
and for the two different $\Delta=0.05, 0.1$.  The results shown below are for
$J_{R}=3$ and $\Delta=0.1$.  The "zero-bias" resistance reported in
experiments may need to be treated carefully.  In fact, the voltage measurment in a
single JJ is limited, unlike 1D or 2D JJ	
arrays, to $V_{min}\sim 0.5\mu V$\cite{Hakonen}.  So we have
calculated the differential resistance 
$
R\equiv\frac{dV}{dI}
$
with the constraint that the voltage is equal to 0.5$\mu V$ (The calculation
is repeated for $V_{min}=0.05\mu V$ to find no essential difference from the
results below).  At the same time, to be
consistent with our theoretical consideration above, 
we have checked that the current is much smaller than the critical one, $I\ll
I_{0}$.  For results below, the current $I$ is up to a few \% of the critical one
$I_{0}$.  
Fig.~1 shows a temperature dependence of the resistance at $V=0.5\mu V$ for
$g=0.4$.  
The small $E_{J}/E_{C}$ case shows an
insulator-like temperature dependence $\frac{dR}{dT}<0$, while the large
$E_{J}/E_{C}$ case shows a superconductor-like behavior $\frac{dR}{dT}>0$.
 Repeating the calculations for different g, one can draw a phase diagram.  Fig.~2 is
a phase diagram at $T=80~mK$.  The experimental results\cite{Hakonen,Yagi}
are open circles (superconductor-like) and solid circles (insulator-like) at
$T\approx 80~mK$.  Our result is denoted by open diamonds.  The
band theory result, thick solid line, is obtained by simply replacing the
minimum measurament energy $eV_{min}$ by the thermal energy
$k_{B}T$\cite{Hakonen}.  The QTS theory result, triangles, is also plotted for
comparison.
 A major disagreement between theory and experiment
is for some data points near $g=2.8$\cite{Yagi}.  However a similar phase
diagram experimentally found for the 2D JJ arrays with similar
parameter ranges for $g, E_{j}/E_{C}$ and $T$ is bounded, $E_{J}/E_{C}\lesssim
0.5$, and $g\lesssim 0.5$ (cf. Fig~3 in\cite{Yamaguchi}).  The above data near $g=2.8$
is currently mysterious.

There are two notable points in the obtained temperature dependence of
resistance.  First it becomes flat for
large $E_{J}/E_{C}$ irrespective of $g$.  This is consistent with
experimental findings\cite{Hakonen,Yagi}.  Theoretically, when the
ratio $E_{J}/E_{C}$ is sufficiently large, the quantum Kramers rate
 would become less dependent on $T$ and $g$.  Note a
sharp contrast between the single JJ and the 2D JJ
arrays\cite{Yamaguchi}.
The temperature dependence of resistance in the latter case showed a {\it
true} phase transition behavior, namely resistance decreases or increases
roughly {\it
exponentially} with temperature.  Interestingly, the temperature dependence of
resistance in 1D JJ arrays also becomes flat deep inside the superconducting phase\cite{Kuo}.
The second notable point
is that the SI phase boundary defined by vanishing $dR/dT$ actually depends on
temperature.  In Fig.~3, we have plotted the SI phase boundary for
$T=80, 60, 40, 20, 10$ and $5~mK$.  Clearly, the insulator phase diminishes
with decreasing temperature.
We thus reach a conclusion that, {\it as long as dissipation is present, the single JJ
becomes all superconducting at absolute zero temperature}.  Note that,
in contrast, the band theory predicts an essentially temperature independent
phase boundary.  It is interesting
to see if our finding can be observed experimentally.

Finally, related to the last point, we should argue that our finding is indeed consistent with the
previous theoretical works on a different RSJ
model\cite{Gravert,Hanggi}
 and the subsequent experimental
observations\cite{Washburn} of a universal $T^2$-law for the escape rate
$k^+$ of the zero-voltage state in the single JJ.  That is,
\begin{equation} \label{eq17}
ln[k^+(T)/k^+(0)]\propto T^2~~{\rm for}~~T\to 0
\end{equation}
  These experiments focused on the highly dissipative regime $g\gg 1$,
and the current is very close to the critical current, and therefore the
escape rate $k^+$ is not quite proportional to voltage.  In contrast, we focused
on the regime $g\lesssim1$ and calculated voltage since current is only
up to a few \% of the critical one.  Nevertheless the present
theory and the previous theories agree in that both $k^+$ and
$R(T)$ increase with temperature at very low T.  In fact it was argued that
the universal low temperature enhancement arizes from the temperature
dependence of the heat-bath\cite{Hanggi}.  In the present case, it is due to
the $\alpha(\tau)$ term in the action (\ref{eq3}) and its temperature
dependence originates from the Fermi distribution function in the electron
Green's function
\cite{Ambegaokar}.  As $T \to 0$, this term dominates
the temperature dependence and the voltage behaves as 
\begin{equation} \label{eq18}
ln[V(T)/V(0)]\propto gT^2
\end{equation}
leading to $\frac{dR}{dT}>0$ for $T\to 0$.  Note that, as is clear from (\ref{eq18}) and seen in
Fig.~3, such a temperature
dependence disappears at $g=0$. 

This work was supported by the Visitor Program of the MPI-PKS. The work was also partially supported by the NSF under Grant 
No. DMR990002N and utilized the SGI/CRAY Origin2000 at the National Center 
for Supercomputing Applications at the University of Illinois at
Urbana-Champaign. I thank Mikhail Fistul, Pertti Hakonen and Pavel
Lipavsk$\acute{{\rm y}}$ for helpful discussions and Chiidon Chen and Watson Kuo 
for correspondence on their experiment.  I also thank Peter H\"anggi, Ivan
Larkin and Anthony Leggett for enlightening conversations.


\begin{figure} \label{fig04}
\epsfig{file=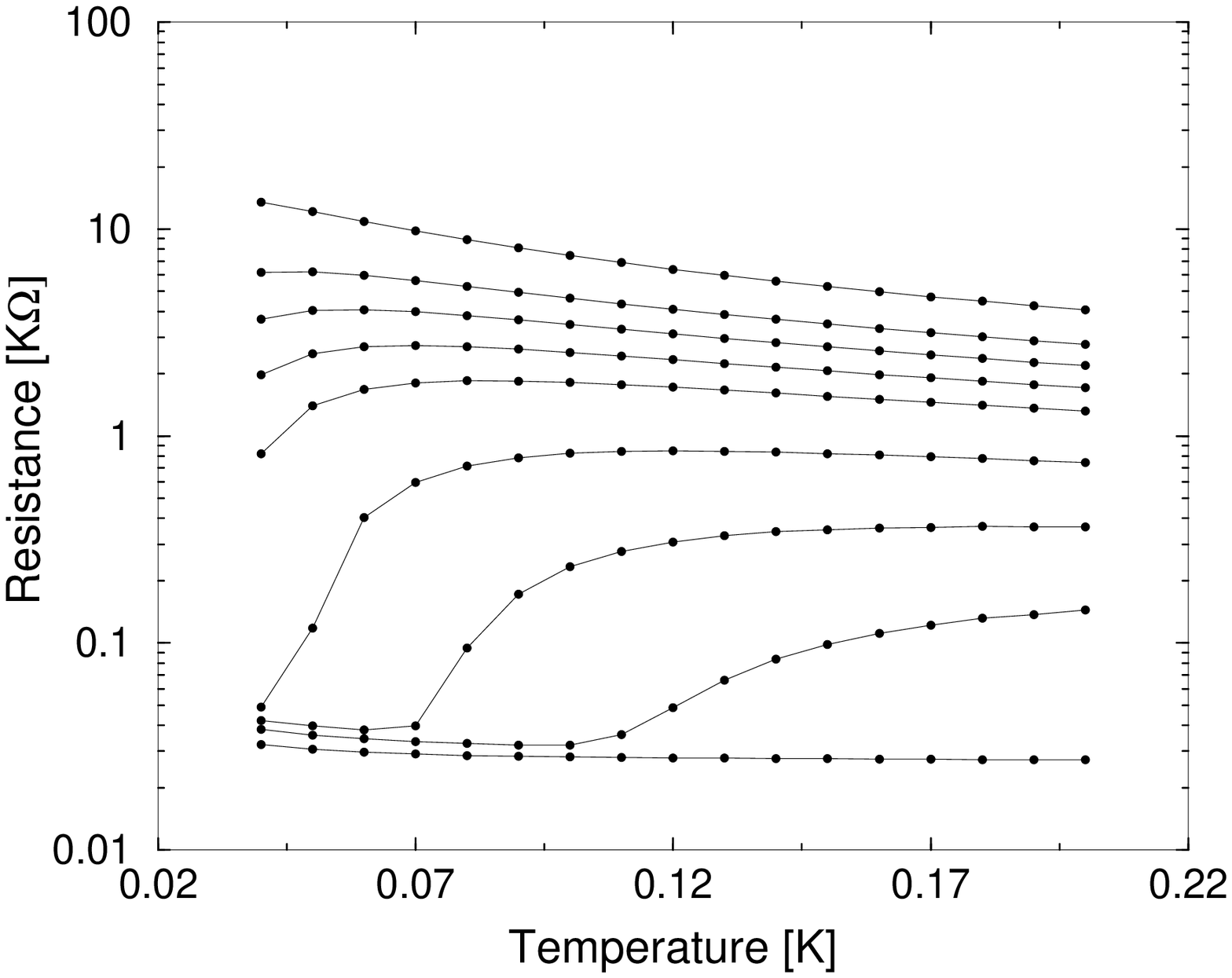,width=7cm,height=7cm,angle=-90}
\end{figure}
\vspace{-0.5cm}
FIG.~1. Temperature dependence of the "zero-bias" resistance
at $g=0.4$. From top to bottom, $E_{J}/E_{C}$=0.07, 0.56, 0.84,
1.12, 1.4, 1.96, 2.52, 3.0 and 4.0.

\begin{figure} \label{jjpost_4}
\epsfig{file=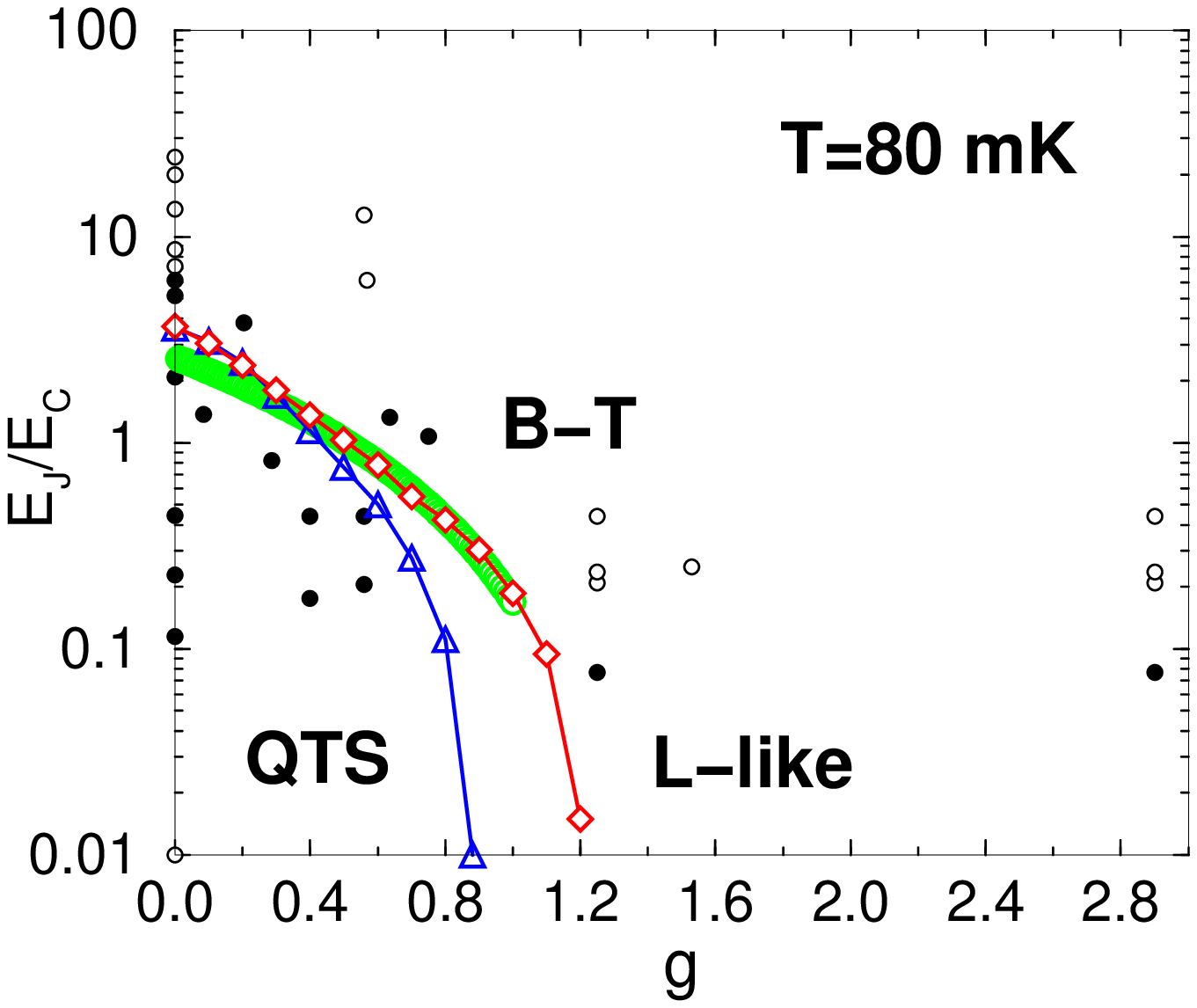,width=7cm,height=7cm}
\end{figure}
\vspace{-0.5cm}
FIG.~2. Phase diagram of shunted Josephson junction at $T=80mK$. 
The phase boundary lies
between the insulator-like (solid circles) and superconductor-like (open
circles) samples experimentally found in [7,8].  The thick line
is the band theory(B-T), the triangle the QTS theory and the
diamond is due to the present theory(L-like).
\begin{figure} \label{phase5}
\epsfig{file=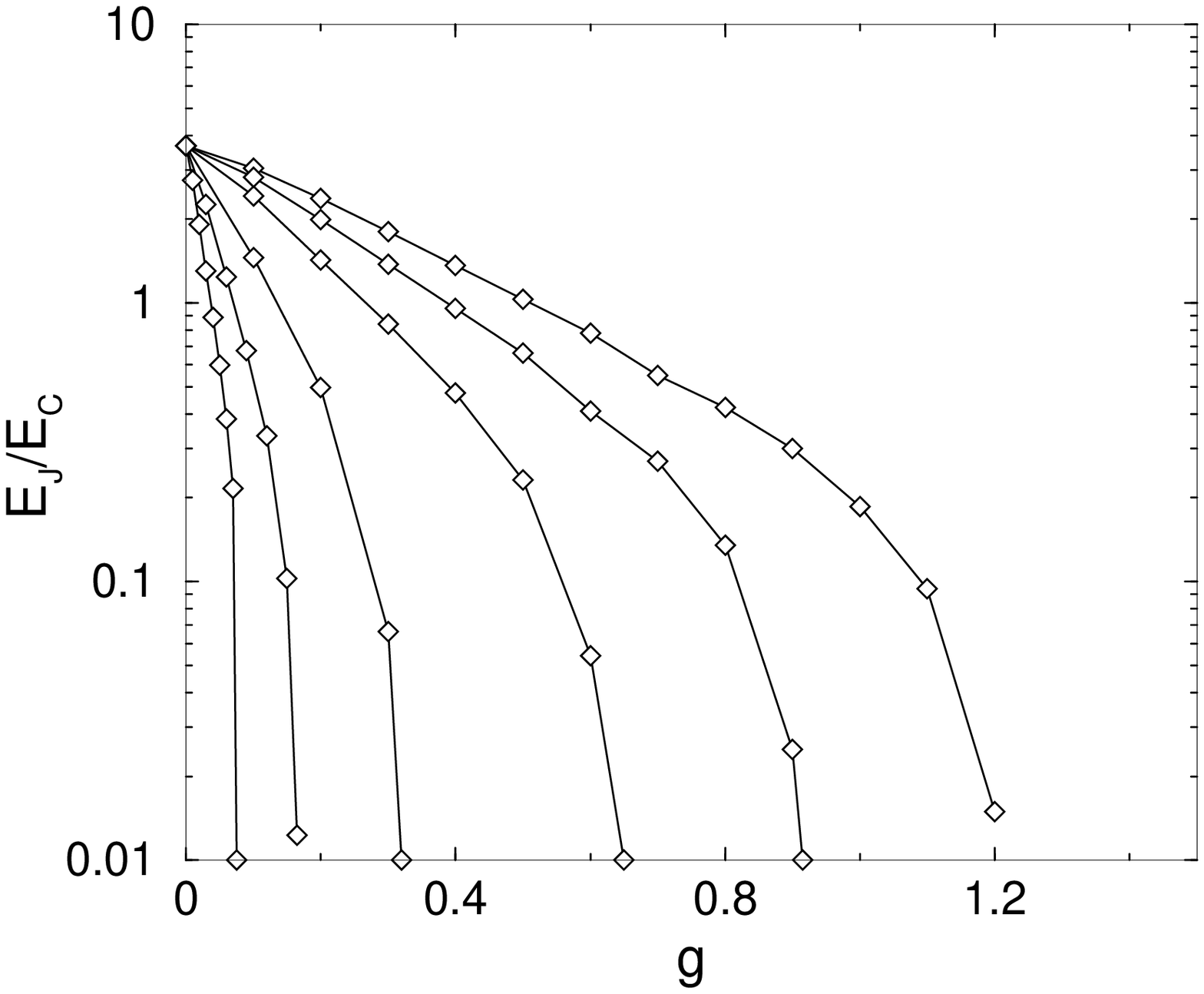,width=7cm,height=7cm,angle=-90}
\end{figure}
\vspace{-0.5cm}
FIG.~3. Temperature dependence of the superconductor-insulator phase boundary
in the $E_{J}/E_{C}-g$ plane. From right to left, $T=80, 60, 40, 20, 
10,$ and $5~mK$.

\end{multicols}

\end{document}